\documentclass[preprint, 12pt]{aastex63} 

\usepackage[T1]{fontenc}

\usepackage{bm}        
\usepackage{amssymb, amsmath}   
\usepackage{cancel}
\usepackage{color}
\usepackage{url}

\usepackage[utf8]{inputenc}
\usepackage[english]{babel}

\usepackage{graphicx}
\usepackage{float}
\usepackage{amsmath, amsthm, amsfonts}
\usepackage{tcolorbox}
\usepackage[verbose]{wrapfig}
\usepackage{color}
\usepackage{url}
\usepackage{natbib}
\usepackage{color}
\usepackage{transparent}

\definecolor{tdblue}{HTML}{204A87}
\definecolor{tmblue}{HTML}{3465A4}
\definecolor{tlblue}{HTML}{729FCF}
\usepackage{hyperref}

\begin{document}

\title[Shock Propagation and Particle Acceleration in Solar Wind Turbulence]{Shock Propagation and Associated Particle Acceleration in the Presence of Ambient Solar-Wind Turbulence}

\author{Fan Guo}
\affiliation{Theoretical Division, Los Alamos National Laboratory, Los Alamos, NM 87545, USA}
\affiliation{New Mexico Consortium, Los Alamos, NM 87544, USA}
\author{Joe Giacalone}
\affiliation{Lunar \& Planetary Laboratory, University of Arizona, Tucson, AZ 85721, USA}
\author{Lulu Zhao}
\affiliation{Department of Climate and Space Sciences and Engineering, University of Michigan, Ann Arbor, MI, 48109, USA}



\date{\today}

\begin{abstract}
\noindent The topic of this review paper is on the influence of solar wind turbulence on shock propagation and its consequence on the acceleration and transport of energetic particles at shocks. As the interplanetary shocks sweep through the turbulent solar wind, the shock surfaces fluctuate and ripple in a range of different scales. We discuss particle acceleration at rippled shocks in the presence of ambient solar-wind turbulence. This strongly affects particle acceleration and transport of energetic particles (both ions and electrons) at shock fronts. In particular, we point out that the effects of upstream turbulence is critical for understanding the variability of energetic particles at shocks.  Moreover, the presence of pre-existing upstream turbulence significantly enhances the trapping near the shock of low-energy charged particles, including near the thermal energy of the incident plasma, even when the shock propagates normal to the average magnetic field. Pre-existing turbulence, always present in space plasmas, provides a means for the efficient acceleration of low-energy particles and overcoming the well known injection problem at shocks.

\end{abstract}

\section{Introduction}

Understanding shocks and its particle acceleration in the solar wind is an important topic in heliophysics, for both the effects of space weather and basic physics of shocks and particle energization. Near the Sun, the shocks emerge in front of the solar drivers and accelerate solar energetic particles (SEPs) \citep{Reames1999,Desai2016}. However, those shocks are farther away from most of the spacecraft, making them difficult to observe directly. At 1 astronomical unit (AU), these shocks are mostly driven by coronal mass ejections, with a smaller population driven by Stream Interaction Regions (SIRs) \citep{Sheeley1985,Richardson2010,Borovsky2020}. Strong energetic particles observed in the vicinity of interplanetary shocks are often termed as ``Energetic Storm Particles'' (ESP) \citep{Bryant1962,Gosling1981}, to distinguish them from the solar energetic particles (SEPs) accelerated close to the Sun, although the SEPs can be well accelerated by the coronal counterpart of the interplanetary shock driven by the same coronal mass ejection \citep{Kallenrode1996,Reames1999}. Most interplanetary shocks are weak compared to planetary bow shocks. Because of the Parker spiral magnetic field, the interplanetary shocks at 1 AU are mostly quasi-perpendicular, meaning the angle between the un-shocked, upstream magnetic field and the unit normal to the shock $\theta_{Bn}$ is usually larger than $45$ degree. 
Decades of spacecraft observations by Helios, IMP, ISEE, ACE, Wind, STEREO and others have accumulated tremendous amount of shock events (as many as several dozens per year at solar maximum) for study \citep{Neugebauer2013,Dresing2016}, including two widely used shock lists collected by ACE \footnote{ACE shock list: \url{http://www.ssg.sr.unh.edu/mag/ace/ACElists/obs_list.html}} and Wind \footnote{Wind shock list: \url{https://www.cfa.harvard.edu/shocks/}}. Newly launched Parker Solar Probe and Solar Orbiter provide shock and energetic particle observations close to the Sun, and will further reveal the physics of interplanetary shocks and their associated energetic particles. The upcoming IMAP mission will provide high resolution measurements of suprathermal ions and provide further observation insight, especially on the variability of energetic particles in the vicinity of the shocks \citep{McComas2018}.

In addition to the mean, Parker spiral magnetic field, the solar wind is well known to be filled with numerous magnetic structures \citep{Borovsky2008,Neugebauer2010} and turbulence \citep{Tu1995,Goldstein1995}. In situ measurements of the solar wind have long established the existence of turbulence that has a Kolmogorov-like power spectrum with a correlation scale of about $10^6$ km at 1 AU and increases in the outer heliosphere, but could also be contributed by magnetic structures and discontinuities \citep{Borovsky2010}. Unlike Earth's bow shock that is on a much smaller spatial scale and more suitable for kinetic studies, the interplanetary shocks provide a natural laboratory for physical processes involving high energy charged particles and large scale magnetic field fluctuations. Compared to other shock-turbulence systems like the solar wind termination shock and supernova remnant shocks, interplanetary shocks are frequently observed, with a wealth amount of data accumulated.  The propagation and evolution of interplanetary shocks in the solar wind are influenced by fluctuations in magnetic fields, velocity, and density. The turbulent magnetic field can interact with the shock waves, distorting their surfaces, leading to shock ripples \citep{Neugebauer2005} and enhance the downstream magnetic fluctuations \citep{Zank2003,Lu2009}. It is also important for efficient particle acceleration \citep{Giacalone2005,Jokipii2007,Guo2010Particle,Guo2010Effect,Guo2015Acceleration}. These nonplanar ripples at the shock surface, coupled with turbulent upstream magnetic fields, lead to substantial variations of energetic particle flux, which has been pointed out, but never been understood in a systematic way \citep{Giacalone2008,Guo2010Particle,Kota2010}.

The diffusive shock acceleration (DSA) theory describes the basic process of acceleration of particles at the shock front. A review of the mechanism is given by \citet{Drury1983} (see also \citet{Desai2016}), and was independently discovered by \citep{Krymsky1977,Axford1977,Bell1978,Blandford1978}. For the often considered one-dimensional solution, this gives the classical results:

\begin{eqnarray} 
f_s(p) = f_0 p^{-3r/(r-1)} H(p - p_0)
    \begin{cases} \hspace{5pt}
      \exp(x U_1 / \kappa) & x<0 \\
      1 & x \geq 0  \end{cases}
\label{eq:DSASoln}
\end{eqnarray}

\noindent where $f_0$ is a normalization constant, $r$ is the ratio of the downstream to upstream density, $U_1$ is the upstream flow speed in the shock frame, and $H(p)$ is the Heaviside step function. This solution is obtained by solving the Parker transport equation (see a discussion in Section \ref{sec:Injection}) for a one-dimensional time-steady shock at $x=0$.

However, this result has only received limited success when applied to the interplanetary shocks. While there has been claims that an excellent agreement can be made between observations \citep{Kennel1986} and theory \citep{Lee1983} for an interplanetary shock event measured on November 12, 1978, even for that particular event, the observation over a longer time scale shows more complicated variation \citep{Scholer1983}. In fact, most of energetic particle profiles in the vicinity of interplanetary shocks are not consistent with the 1-D solution \citep{Lario2003}. 
While in some cases particles that interact with interplanetary shocks may have been accelerated in solar events \citep{Li2005}, this variation can also be induced by the fact that the 1-D solution fails due to the large-scale magnetic fluctuation, even without previous events \citep[e.g.,][]{Neugebauer2006,Giacalone2008}.
We will discuss the modification of the shock front due to the solar wind turbulence in Section \ref{Sec:NonplanarShock}. The pre-existing fluctuations are particularly important for the acceleration of particles at quasi-perpendicular shocks. 
We will discuss the effects of solar wind turbulence on the acceleration of protons and electrons at shocks in Section \ref{sec:Injection} and \ref{sec:Electron}, respectively.

Understanding shock-turbulence interaction and particle acceleration in the turbulent solar wind also has strong implication to other turbulence system, such as the magnetic field amplification when supernova shocks sweeping through the interstellar medium density and the roles of magnetic field fluctuations on particle acceleration \citep[e.g.,][]{Giacalone2007,Inoue2009,Guo2012amplification,Fraschetti2013}. 

Rest of this paper is organized as follows: We discuss the nonplanar shock effects led by the solar wind turbulence in Section \ref{Sec:NonplanarShock}. A discussion on the acceleration of protons is given in Section \ref{sec:Injection}, focusing on the injection problem of protons. We discuss the issue about electron acceleration at interplanetary shock in Section \ref{sec:Electron}. We emphasize the variability caused by turbulence in Section \ref{sec:Variability}. We leave some final remarks in Section \ref{sec:final}.

\section{Nonplanar shocks led by turbulence}
\label{Sec:NonplanarShock}

\begin{figure}[h!]
\begin{center}
\includegraphics[width=10cm]{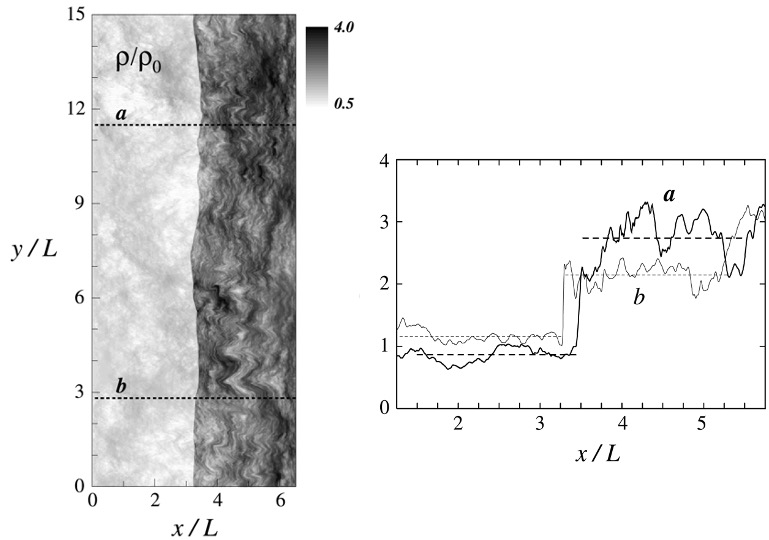}
\end{center}
\caption{Left panel shows a gray-scale representation of simulated plasma density from a  two-dimensional MHD simulation. $L$ is the correlation scale of the turbulent fluctuations. The shock is rippled as it swept through the turbulent upstream plasma. Right panel shows two cuts for density profiles at different y positions indicated on the left panel (adapted from \citet{Giacalone2008}).}\label{fig:rippledshock}
\end{figure}

As interplanetary shocks sweep through the turbulent solar wind, their surfaces naturally distort and fluctuate on a variety of scales. This fluctuation in the warped shock surfaces occurs not only in the large scale due to the large-scale turbulent plasma and magnetic fields \citep{Zank2003,Giacalone2005,Li2006,Giacalone2008,Guo2010Effect,Guo2012Acceleration,Trotta2020}, but at small scale as well, due to instabilities caused by ion reflection \citep{Lowe2003,Burgess2006,Yang2012,Hao2016,Trotta2019}. Using hybrid kinetic simulations (fluid electrons and kinetic ions) and MHD simulations, \citet{Giacalone2005}, \citet{Giacalone2008} and \citet{Guo2010Effect} have shown that the shock surface is wrapped due to the pre-existing upstream magnetic fluctuations. The angle between the shock normal and the incident magnetic field, $\theta_{Bn}$, vary along the shock surface. Figure \ref{fig:rippledshock} shows plasma density in a 2D MHD simulations where the shock propagates through a fluctuating plasma. It clearly shows that the shock is rippled in various scales from that comparable to the system size down to very small scales. The density profiles can be quite variable as indicated by two cuts across the shock.

The idea that interplanetary shock surface is wrapped in turbulence has been tested in the solar wind by \citet{Neugebauer2005}. They analysed 26 well-defined quasi-perpendicular
interplanetary shocks that were observed by five spacecraft or more. They used the four spacecraft
method to obtain five (five in the six spacecraft event) independent
sets of normals and speeds. They also calculated normals and
speeds using single-spacecraft methods. Furthermore, they estimated the
shock radius of curvature with different methods. Most of the shocks were
inconsistent with planar structures, or spherical structures with a radius of
1AU. In other words, the shocks were found to be rippled. They reported
that the average local radius of curvature is $\sim 3 \times 10^6$ km ($\sim 500R_E$)
which is close to the observed correlation scale of the fluctuating IMF. Note
that the size of the ripples they could observe was limited from below by
the spacecraft separations, which were of the order of $10^5$ km, corresponding
to $10^3$ ion inertial lengths for typical solar wind conditions. Thus they
were not able to discern any smaller scale (a few times of the ion inertial length) ripples existing at the same time, though such were evident in the 2D hybrid
simulations presented in the same article. Later \citet{Koval2010} confirmed this basic conclusion. The scale of measured shock ripples is similar to the correlation length of the interplanetary
turbulence \citep[e.g.,][]{Coleman1968,Matthaeus1982,Chen2012}. In a recent study, irregular surface of interplanetary shocks at ion scales has been verified by multispacecraft observations of the Cluster mission \citep{Kajdic2019}.

The shock ripples and large-scale magnetic fluctuations have also been invoked to explain the variability of interplanetary type II radio bursts \citep{Bale1999,Pulupa2008,Knock2003}. \citet{Bale1999} and \citet{Pulupa2008} observed electron foreshocks and
related Langmuir waves upstream of interplanetary shocks. These events appeared
to be associated with irregular shock surfaces with spatial scales $ \sim 10^5$ km, in rough consistency with \citet{Neugebauer2005}.
They proposed that the complex upstream electron events result from large-scale
irregularities in the shock surface. 


\section{The Physics on the Injection Problem and its Importance to Understanding the Source of Solar Energetic Particles}
\label{sec:Injection}

Pre-existing turbulence also has an important effect on low-energy charged particles, including so-called suprathermal particles whose distribution function connects smoothly to the thermal part of the incident plasma distribution. In order to understand this, it is important to first discuss the physics of particle acceleration more generally. This is directly related to the understanding, and determination of the source of high-energy particles accelerated at shocks. It is also critical to understand the so-called `injection problem' at shocks. 

The mechanism for DSA has been discussed in many dedicated papers mentioned in Section 1, and it is generally accepted that this mechanism applies above some `injection' energy, $E_i$. In this section we focus on the question of what happens at energies below $E_i$, and how it relates to the intensity and shape of the spectrum above this energy. 

On the one hand, stated simply, the theory of diffusive shock acceleration does not address the injection process; thus, it might be easiest to just assume that there exists a pre-existing population of particles and that these are the particles which are further accelerated by shocks ~\citep[e.g.,][]{Mewaldt_etal_2012}. Moreover, the largest energetic particle events associated with coronal mass ejections often come from large active regions on the Sun which commonly also produce a number of smaller CMEs and transient shocks. It is reasonable to expect that these smaller events produce energetic particles that could be accelerated further at the later shock (c.f. \citet{Desai2016}, for a review of observations). It is not clear, however, whether it is the existence of the pre-existing population of energetic particles, or the fact that the later shock moves through a more (magnetically) turbulent medium, which increases the trapping of the particles at the later shock, leading to a higher acceleration rate, which causes the higher fluxes of particles at very high energies.

On the other hand, it is important to examine the physics of particle acceleration at low energies, and to determine the conditions under which DSA theory is applicable. DSA is based on the Parker transport equation, which is given by Equation \ref{eq:ParkerEquation}, and was first derived by \citet{Parker1965}. This equation assumes the particle's distribution is nearly isotropic. 

\vspace{5pt}
\subsection{The Proper Frame of Reference for the Energy-Change Term in the Parker Transport Equation}
\label{subsec:EchangeTrm}

We start with an analysis of the term that is responsible for the acceleration, paying particular attention to the proper frame of reference since the pitch-angle distribution of low-energy particles varies considerably between the shock frame and the plasma frame. 

We now show that this term follows directly from the basic physics of charged (test) particle motion in electric and magnetic fields. We follow closely the approach of \citet{Jokipii2012}, which has rarely been cited in the literature, suggesting that this important derivation has largely been forgotten.  We add some more steps to this derivation which did not appear in the original manuscript, for completeness.

The motion of a charged particle, with charge $q$, in a plasma containing an electric field $\textbf{E}$ and magnetic field $\textbf{B}$ is given by:

\begin{eqnarray} 
\frac{d \textbf{p}}{dt} = q \textbf{E} + \frac{q}{c} \textbf{w} \times \textbf{B}
\label{eq:LorentzForce}
\end{eqnarray}

\noindent where \textbf{p} is the momentum of the particle, \textbf{w} is the velocity vector, and c is the speed of light. Gravity and radiation pressure, two forces which are commonly included in macroscopic equations because these forces are negligible in most situations of interest regarding the acceleration of particles.

The rate of change of the kinetic energy, $K$,  of a charged particle is obtained by taking the vector dot product of the particle velocity with Equation \ref{eq:LorentzForce}, and is given by:

\begin{eqnarray} 
\frac{dK}{dt} = q \textbf{w} \cdot \textbf{E}
\label{eq:dKdt1}
\end{eqnarray}

We consider the case in which the electric field is given by that of ideal magnetohydrodynamics: $\textbf{E} = -\textbf{U} \times \textbf{B}/c$, where $\textbf{U}$ is the velocity of the bulk plasma which carries with it the magnetic field $\textbf{B}$. Thus, we have

\begin{eqnarray}  \frac{dK}{dt} = - \frac{q}{c} \textbf{w} \cdot (\textbf{U} \times \textbf{B})
\label{eq:dKdt2}
\end{eqnarray} 

We next consider the kinetic energy in the frame moving with the bulk plasma. Using a prime to denote quantities in the plasma frame of reference, we have

\begin{eqnarray}  K' = K - \textbf{p} \cdot \textbf{U} + \frac{1}{2}mU^2
\label{eq:KprimeToK}
\end{eqnarray} 

Inserting this into Equation \ref{eq:dKdt2}, and using Equation \ref{eq:LorentzForce} and a vector identity, after some simple manipulation it follows that:

\begin{eqnarray} 
\frac{dK'}{dt} = -(\textbf{p} - m\textbf{U}) \cdot \frac{d \textbf{U}}{dt} = - \textbf{p}' \cdot \frac{d\textbf{U}}{dt} = -\textbf{p}' \cdot (\frac{\partial \textbf{U}}{\partial t} + \textbf{w}'\cdot \nabla \textbf{U})
\label{eq:dKpdt}
\end{eqnarray} 

\noindent where the middle term on the right is from the definition of the momentum in the plasma frame, and the last term on the right follows by noting that the total time derivative is that along the particle trajectory in which the plasma velocity can change both temporally and spatially; hence, the use of the advective derivative is appropriate.

We note that the kinetic energy depends on the vector momentum and velocity, and, thus not only the magnitude of these quantities but also their direction: i.e. the particle pitch angle and phase angle. We now consider a distribution of particles and average Equation \ref{eq:dKpdt} over pitch and phase angle. Denoting this average with the angle-bracket notation, we have

\begin{eqnarray} 
\frac{d \left<K' \right>}{dt} = \left<-\textbf{p}' \cdot (\frac{\partial \textbf{U}}{\partial t} + \textbf{w}'\cdot \nabla \textbf{U}) \right>
\label{eq:dAVGKdt1}
\end{eqnarray} 

It is important to note that the average is taken over the plasma-frame pitch-angle distribution. We assume that the magnetic fluctuations will `scatter' the particles and alter particles' pitch angles \citep{Jokipii1971} with their energy conserved upon each scattering in the plasma frame. If the time between scatterings is shorter than the time variation of the plasma velocity, then the first term inside the parenthesis on the right of Equation \ref{eq:dAVGKdt1} is small and can be neglected. Switching to index notation for the vectors, we have:

\begin{eqnarray} 
\frac{d \left<K' \right>}{dt} = -\left< p'_i \cdot w'_j \right> \partial_j U_i
\label{eq:dAVGKdt2}
\end{eqnarray} 

In deriving this expression, one must also assert that the variation in space of the fluid velocity is on a scale larger than the scattering mean-free path of the particles, which is simply the magnitude of the particle velocity times the scattering time. Assuming that the distribution is isotropic, the term inside the angle brackets in this equation only gives an answer when $i=j$ and also gives a factor of $1/3$ times the magnitude of the momentum and velocity. Equation \ref{eq:dAVGKdt2} also permits contributions from fluid shear \citep{Parker1965,Earl1988, Li2018,Du2018}.

Noting also that the kinetic energy can be written in terms of the momentum ($dK’/dt=w’dp’/dt$), Equation \ref{eq:dAVGKdt2}, in terms of the rate of change of the momentum in the plasma frame for a nearly isotropic distribution, is given by:

\begin{eqnarray} 
\frac{d \left<p' \right>}{dt} = -\frac{1}{3} p' \nabla \cdot \textbf{U}
\label{eq:dAVGpdt}
\end{eqnarray} 

This represents the rate of change of the magnitude of the momentum, in the {\it plasma frame} of reference, in a plasma with bulk velocity \textbf{U}. We note that energy change only occurs when the plasma has a finite divergence, either rarefactions or compressions of the plasma. In the case of a shock wave, the plasma velocity decreases across the shock, in the shock frame of reference, so that the divergence is negative. Thus, the plasma-frame momentum of the particles increases across the shock, and the energy change occurs directly at the shock.  This will be discussed further below.

In the cosmic-ray (or Parker) transport equation first derived by \citet{Parker1965}, given by Equation \ref{eq:ParkerEquation} below, the term representing the energy change of an isotropic distribution of particles is of the same form as Equation \ref{eq:dAVGpdt}. Thus, in the Parker equation, the momentum variable is that measured in the plasma frame of reference. The Parker equation is given by:

\begin{eqnarray} 
\frac{\partial f}{\partial t} + \textbf{U} \cdot \nabla f = \nabla \cdot (\kappa \nabla f) + \frac{1}{3} \nabla \cdot \textbf{U} \frac{\partial f}{\partial \text{ln} p} + S - L
\label{eq:ParkerEquation}
\end{eqnarray} 

\noindent where the first term on the right represents spatial diffusion, and is discussed further below. The second term on the right is energy change, as discussed above, while the last two terms of the Parker equation represent sources and losses.

The spatial diffusion tensor appearing in the Parker equation depends on the magnetic field, both its turbulent, fluctuating component, as well as its average component. It can be decomposed into components along and across the average magnetic field, and a term which represents drift motions associated with gradient and curvature drifts, according to:

\begin{eqnarray} 
\kappa = \kappa_{ij} = \kappa_\perp \delta_{ij} + (\kappa_\parallel - \kappa_\perp) \frac{B_iB_j}{B^2} + \epsilon_{ijk}\kappa_A \frac{B_k}{B}
\end{eqnarray}

\noindent where the vector $B_i$ represents the average magnetic field. $\kappa_\perp$ and $\kappa_\parallel$ are the diffusion coefficients across and along the magnetic field and these coefficients depend on rigidity of the particles as well as the turbulent component of the magnetic field \citep{Jokipii1971,Giacalone1999}.  $\kappa_A$ represents particle drifts, including those associated with the curvature and gradient of the average magnetic field \citep{Isenberg1979}.

It is important to note here that the only assumption is that the distribution is isotropic, in the local plasma frame of reference. Above, when we derived the form of the energy-change term, we noted that the momentum was that in the plasma frame. Of course, for very high energy particles, the particle speed is many times greater than the bulk plasma speed, thus, the distinction between the plasma and inertial frames is not much. However, it is important in the context of understanding the acceleration at low energies, and the injection problem, since we are particularly interested in the lowest energy to which the diffusive shock acceleration theory is applicable.

\vspace{8pt}
\subsection{The Injection Problem: The Perspective from the Diffusive Shock Acceleration Theory}
\label{subsec:InjProb}

In this review we are particularly interested in the injection problem of energetic particles at shock waves with different shock angles  $\theta_{Bn}$. Without loss of generality, we can take the magnetic field to be in the $x-z$ plane, and the unit normal to the shock in the $–x$ direction, which is also common. Thus, the component of the symmetric diffusion tensor normal to the shock, for example, can be written in terms of $\theta_{Bn}$ as:

\begin{eqnarray} 
\kappa_{xx} = \kappa_\perp \sin^2 \theta_{Bn} + \kappa_\parallel \cos^2\theta_{Bn}
\label{eq:Kxx}
\end{eqnarray}

It is straightforward to solve the Parker equation for a shock-like discontinuity and the solution is given by Equation \ref{eq:DSASoln}. We note that the momentum, or energy, distribution in the shocked plasma has a power-law dependence on momentum (or energy), with an index that is only a function of the ratio of the upstream to downstream plasma velocity across the shocks. This is also the ratio of the downstream to upstream plasma density, or the shock density jump factor. This result is well known, and the physics is well studied. However, the Parker equation is a purely test-particle treatment and must assume an initial energy of the particles. It does not address where the particles at this initial energy come from, or their intensity and contribution to the overall energetics of the entire plasma. It is therefore instructive to consider the limits of applicability of the Parker equation in order to understand this particular issue better.

The Parker equation is derived by assuming the distribution is nearly isotropic in pitch and phase angle. Thus, it is averaged over these quantities, and neither appear in the equation. Thus, it is implicitly assumed that the distribution is nearly isotropic. Yet, because the equation has a diffusive term, there exists a diffusive anisotropy. In order for this equation to be internally consistent, this diffusive anisotropy must be small. At a shock, the distribution at any given energy, upstream of the shock, decays exponentially from the shock towards the upstream region (see our derivation below). The gradient in the particle intensity, in this case is purely diffusive, and leads to a diffusive streaming flux along the shock-normal direction, $S_x$, of the form: $S_x = -\kappa_{xx} \nabla f$. The normalized diffusive anisotropy is $\delta = 3S_x/(wf)$, where $w$ is the particle speed. The diffusive length scale associated with the exponential decay of particles away from the shock in the upstream region is $\kappa_{xx}/U_1$ (c.f. Equation \ref{eq:DSASoln}), where $U_1$ is the upstream plasma velocity, so that by using Equation \ref{eq:KprimeToK} in our expression for $S_x$, requiring $\delta \ll 1$, and after some manipulation, one obtains the following constraint on the applicability of the Parker equation applied to a shock:

\begin{eqnarray} 
w \gg 3U_1 \left[ 1 + \frac{\eta^2 \sin^2 \theta_{Bn} + (1 + \epsilon^2) \sin^2 \theta_{Bn} \cos^2 \theta_{Bn}}{(\epsilon \sin^2 \theta_{Bn} + \cos^2 \theta_{Bn})^2} \right]^{1/2}
\label{eq:injvelocity}
\end{eqnarray}

\noindent where $\epsilon = \kappa_\perp / \kappa_\parallel$  and $\eta = \kappa_A / \kappa_\parallel$. This expression was first derived by \citet{Giacalone1999} and \citet{Zank2006}.

The smallest possible value of the second term inside the brackets (inside the radical) is zero. It cannot be negative. Thus, for any value of $\theta_{Bn}$, the Parker equation is only valid for $w \gg 3U_1$. Defining the sonic Mach number, $M$, as the ratio of the upstream flow speed to the upstream sound speed, we have: $w \gg 3Mv_{th,1}$. Since the most intense energetic particle events are presumably associated with shocks of fairly high Mach number, this is a particularly stringent requirement.  This is true equally for both parallel ($\theta_{Bn}=0^\circ$) and perpendicular shocks ($\theta_{Bn}=90^\circ$).  

Under certain conditions, the injection speed given by Equation \ref{eq:injvelocity} has a very strong dependence on the shock-normal angle.  For $\eta \ll 1$ and $\epsilon \ll 1$, this gives $w \approx 3U_1 \sec^2 \theta_{Bn}$, which is the (3x the) speed of a particle moving with the intersection point of a (perfectly straight) field line and the shock front. This leads to the notion of an injection threshold problem for nearly perpendicular shocks since $w$ is very large as $\theta_{Bn} = 90^\circ$. Some hybrid simulations, which satisfy these conditions, confirm this \citep{Giacalone2000,Caprioli2014}. These conditions are not realistic for a wide variety of heliospheric and astrophysical shocks, however. The conditions noted above unrealistically neglects particle motion normal to the average magnetic field. In the limit $\theta_{Bn} = 90^\circ$, Equation \ref{eq:injvelocity} reduces to $w=3U_1(1+\eta^2/\epsilon^2)^{1/2}$, whereas in the limit $\theta_{Bn} =0^\circ$, we get $w=3U_1$. These differ only by the value $(\eta/\epsilon)^2 = (\kappa_A/\kappa_\perp)^2$ which depends on the magnetic field, including its turbulent component and is not necessarily large. Field-line wandering, caused mostly by the largest scales in the fluctuating component of the magnetic field, significantly enhances the motion of particles across the mean magnetic field such that this term can be small.  This is why pre-existing turbulence is fundamentally important to the injection problem. In this case, the acceleration of low-energy particles at a perpendicular shock can be nearly as efficient as that at a parallel shock.  

\begin{figure}[h!]
\begin{center}
\includegraphics[width=\textwidth]{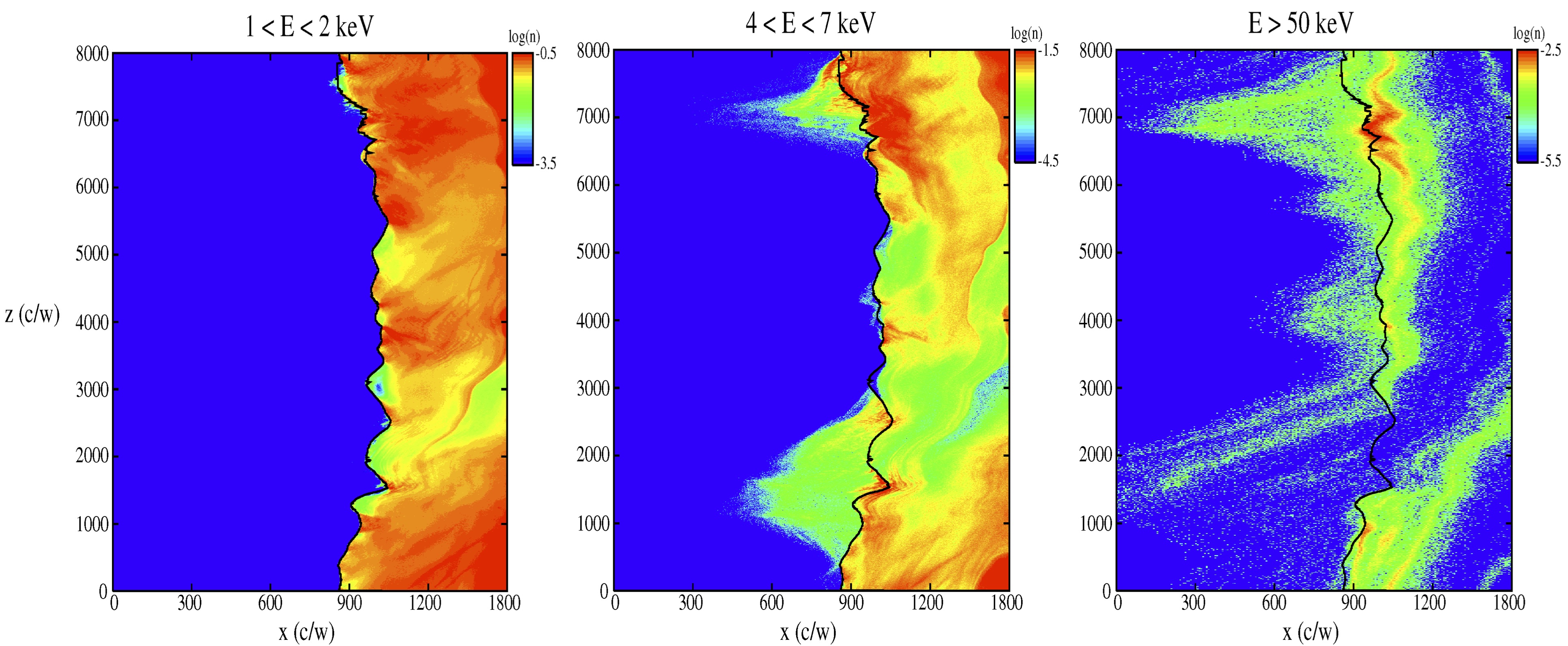}
\end{center}
\caption{A hybrid simulation of a quasi-perpendicular shock ($<\theta_{Bn}>=70^\circ$), with parameters consistent with those of the DOY 118, 2001 interplanetary shock measured by ACE. Shown are suprathermal proton densities at $\Omega_{ci}t=500$ with energies measured in local plasma frame. This result is similar to \citet{Giacalone2005}.}\label{fig:proton-distribution}
\end{figure}

Hybrid simulations of shocks moving through pre-existing broadband magnetic turbulence have revealed that even thermal plasma is efficiently accelerated at a perpendicular shock \citep{Giacalone2005}. Figure \ref{fig:proton-distribution} shows a new hybrid simulation with parameters similar to those of strong, quasi-perpendicular interplanetary shocks. For this case, the Alfven Mach number is 5.8, the total plasma beta is 0.54, and the angle between average magnetic field and the $x$ direction is 70$^\circ$. Other parameters are similar to those presented in \cite{Giacalone2005} (see also \cite{Giacalone2017b}). Similarly, in this simulation, the initial magnetic field is a combination of a mean component and a turbulent component. The turbulent component is based on a power spectrum which contains a range of scales from the size of the box down to the ion inertial length. The large scales leads to field-line meandering, which leads to a variation in the angle between the local magnetic field and the unit normal to the shock, whose surface is rippled, as discussed previously. In several places along the shock front,  the local shock-normal angle is near perpendicular, as was observed for the interplanetary shock on DOY 118, 2001 by the {\it Advanced Composition Explorer} (ACE) \citep{Lario2019}, which had similar parameters to those simulated. The model also includes kinetic processes which lead to heating of the particles across the shock and the formation of a suprathermal tail which are ions accelerated directly from the thermal population. The energy spectra of protons from this particular simulation are shown in Figure \ref{fig:proton-spectrum}.

Observations often show significant enhancements of low-energy ions at quasi-perpendicular interplanetary shocks \citep[e.g.,][]{Zank2006,Giacalone2012,Neergaard-Parker2014} suggesting the injection energy does not depend strongly on the shock-normal angle.  Low-energy ions were also enhanced by over an order of magnitude at the termination shock \citep{Decker2005,Decker2008}, which is a good example of a nearly perpendicular shock. The energy of these ions is far less than the theoretical injection threshold (Eq. 13) for the nearly perpendicular termination shock, which, is readily shown to be several MeV. Moreover, as we discuss below, there are examples of nearly perpendicular interplanetary shocks which also show a very significant enhancement of very low energy ions \citep{Lario2019}. It is a common misconception that nearly perpendicular shocks are unable to accelerate low-energy ions.

\vspace{5pt}
\subsection{The Source Population}
\label{subsec:Seed}

\citet{Neergaard-Parker2012} derived the solution of diffusive shock acceleration for the case of an arbitrary distribution of pre-existing particles, either from previous events or pre-existing in the solar wind, far upstream of the shock, and also included a separate source at the shock itself (c.f. Eq. 6 in their paper). Assuming that the pre-existing population is the only source of particles, we can rewrite their solution for the distribution function downstream of the shock, as:

\begin{eqnarray} 
f_2(p) = \alpha p^{-\alpha} \int p^{\alpha - 1} f_s(p) dp
\label{eq:DSAarbsrc}
\end{eqnarray}

\noindent where $f$ is the distribution function and the subscript $2$ refers to the post-shock region, and the $S$ subscript refers to the pre-existing population of particles. $p$ is the momentum, and $\alpha=3U_1/(U_1-U_2)$, where $U_1$ and $U_2$ are the upstream and downstream bulk plasma speeds as measured in the normal incidence frame (shock rest frame).

The pre-existing source distribution is often described in terms of a kappa distribution. As we have discussed previously, the lowest energy for which the Parker equation is valid, is much higher than $\frac{1}{2}mU_1^2$; thus, only the high-energy part of the kappa distribution is important.  This is simply a power-law distribution. Thus, it is sufficient to assume a pre-existing source function of the form:


\begin{eqnarray} 
f_s(p) = 
    \begin{cases}
      0 & p<p_0 \\
      f_0\left(\frac{p}{p_0} \right)^{-\delta} & p \geq p_0\\
    \end{cases}
\label{eq:SourceDistribution}
\end{eqnarray}

\noindent where $f_0$ is a normalization constant, and $p_0$ can be regarded as the injection momentum. Later we will fit analytic solutions to observed distributions, and it turns out that $p_0$ is a free parameter.  This is discussed further below.

The complete solution for the downstream distribution is obtained by inserting Equation \ref{eq:SourceDistribution} into Equation \ref{eq:DSAarbsrc}, and is given by.

\begin{eqnarray} 
f_2(p) = \frac{\alpha}{\alpha - \delta} f_0 \left[\left(\frac{p}{p_0}\right)^{-\delta} - \left(\frac{p}{p_0}\right)^{-\alpha}\right]
\label{eq:DSAPlawsrc}
\end{eqnarray}

It should be noted that at $p=p_0$, the downstream distribution is zero. This is because the source is upstream of the shock, and because of acceleration at the shock, there are no particles with a plasma-frame momentum $p_0$ downstream of the shock.

In the limit $\delta \rightarrow \infty$, the pre-existing source only has a value at $p_0$ and this expression reduces to $f_2(p)=C (p/p_0)^{-\alpha}$, valid for $p>p_0$, where C is a constant\footnote{Note that if we take the density of the source population to be $n_0$, and assume the source distribution is isotropic, then we find $f_0=n_0(\delta-3)/(4\pi p_0^3)$, which leads to $C=n_0/(4\pi p_0^3)$.}. This is the usual result from diffusive shock acceleration. In particular, the spectral index depends only on the ratio of $U_1$ to $U_2$, which is the plasma density jump across the shock.

However, it is interesting to note that this solution implies that if the pre-existing source has a much harder spectrum than that from the acceleration at the shock (i.e. $\delta \ll \alpha$), the resulting distribution downstream of the shock retains the spectral index of the source spectrum, but that the intensity increases across the shock by the factor $\alpha/(\alpha-\delta)$.  This might be the case when a very weak shock, with a density jump near unity, encounters a pre-existing population of particles with a rather flat energy spectrum. This suggests that weak shocks do not alter the spectral index of the energetic particle distributions compared to the pre-existing population, but rather, simply boost the intensity. Provided, of course, the pre-existing population has sufficiently hard energy spectrum.

In contrast, however, if the pre-existing source has a spectrum which is steeper than that from acceleration at the shock, the shock primarily accelerates particles from the lower-energy portion of the pre-existing distribution, and the shock-accelerated spectrum is consistent with that predicted by diffusive shock acceleration theory.
We also note from Equation \ref{eq:DSAPlawsrc} that the initial momentum, $p_0$, is an important parameter in determining the resulting intensity downstream of the shock. 

It is instructive to consider the special case in which the pre-existing spectrum has a slope such that $\alpha = \delta$.

\begin{eqnarray} 
f_2(p) = \alpha f_0 \left(\frac{p}{p_0} \right)^{-\alpha} \text{ln} \left(\frac{p}{p_0} \right)
\label{eq:PlawSrcSC}
\end{eqnarray}

Thus, the ratio of the shock-accelerated to pre-existing spectrum is $\text{ln}(p/p_0)^\alpha$. At a given momentum (higher than $p_0$), the enhancement factor depends on the injection momentum, $p_0$. Thus, for this case, in which the pre-existing spectrum has a spectral exponent equal to $\alpha$ (the spectral index from simple DSA theory), the injection momentum plays a critical role in determining by what factor the intensity increases. Moreover, the resulting shock-accelerated spectrum is not even a simple power law. We now consider a specific application of this case.

\begin{figure}[h!]
\begin{center}
\includegraphics[width=0.6\textwidth]{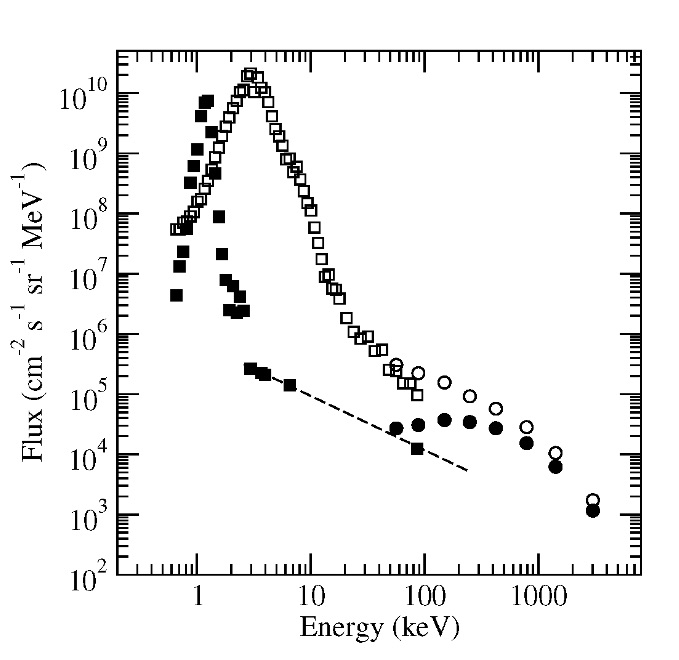}
\end{center}
\caption{Data from \citet{Lario2019}; panels (b) and (f) of their Figure 10. The filled symbols are spectra about 30 minutes prior to the crossing of a shock. The open symbols are about 12 minutes after the shock crossing. The circles are ACE/EPAM data, and the squares are ACE/SWICS data. }\label{fig:lario}
\end{figure}

Figure \ref{fig:lario} is adapted from \citet{Lario2019}. The data shown in this figure is from panels (b) and (f) of their Figure 10, and was provided to us by the authors. The data shows energy spectra upstream and downstream of an interplanetary shock which crossed the spacecraft ACE on DOY 118, 2001. Data from two instruments, SWICS and EPAM, are shown. The original figure, in \citet{Lario2019}, also shows error bars, which are omitted here. The spectra cover a broad range of energy which includes the thermal-peak at low energies, through the suprathermal range, to high energies. Also plotted as a dashed line is the one-count level. as was also done in the original image.

From Figure \ref{fig:lario}, we note that the pre-existing distribution cannot have an intensity larger than the one-count level indicated by the dashed line; thus, at an energy of 250 keV, which is approximately the energy at which the dashed line ends in this figure, the intensity cannot be larger than $5000\ {\rm cm}^{-2} {\rm s}^{-1} {\rm sr}^{-1} {\rm MeV}^{-1}$.   The shock-accelerated particles, downstream of the shock, at this energy have an intensity of $10^5\ {\rm cm}^{-2} {\rm s}^{-1} {\rm sr}^{-1} {\rm MeV}^{-1}$ (open circle at $E = 250$ keV). Thus, the ratio of the shock-accelerated to pre-existing intensity cannot be smaller than 20 ($=10^5/5000$).  Moreover, since the spectral index of the one-count curve in Figure \ref{fig:lario} is approximately the same as that of the shock-accelerated spectrum in the vicinity of 250 keV, Equation \ref{eq:PlawSrcSC} above can be used to determine the value of the injection energy $E_0$. Note also that the ratio of the distribution functions upstream and downstream are the same as the ratio of the differential intensity, in this case, because we are taking the ratio at the same energy. Thus, setting $\text{ln}(p/p_0)^\alpha$ = 20, or, in terms of energy, $\text{ln}(E/E_0)^{\alpha/2} = 20$, we obtain $E_0=E\exp(-40/\alpha)$. $\alpha$ is the spectral index of the distribution function vs. momentum, which can be readily shown to be $2(\gamma+1)$, where $\gamma$ is the spectral exponent of the differential intensity vs. energy, as shown in Figure \ref{fig:lario}. By inspection of Figure \ref{fig:lario}, we find $\gamma \approx 1.1$, and $\alpha=4.2$, so that $E_0=(250\  {\rm keV})\exp(-40/4.2) \approx 0.02$ keV. This is well below the so-called ram energy of the shock (the kinetic energy of the plasma as it enters the shock), and is near that of the thermal energy of the solar wind. 

From the above considerations, we conclude that is not possible to explain the shock-accelerated distribution in this event simply by the ‘lifting up’ of the pre-existing distribution. The intensity of the shock-accelerated particles indicates that the particles were accelerated from a much-lower energy source; and the most-abundant source is the solar wind.  We conclude that for this event, the source of the accelerated particles is the solar wind.

It is also noteworthy that the reported shock parameters for this particular event (Table 1 of \citet{Lario2019}) gives a shock-normal angle of $\theta_{Bn}=88 \pm 2$. This is a very nearly perpendicular shock. Thus, this event shows a clear example of a nearly perpendicular shock which locally accelerates particles from very low energies, forming a high-energy rail. This is consistent with analytic theory and self-consistent plasma simulations which include pre-existing large-scale magnetic fluctuations with parameters consistent with those observed in the solar wind \citep[Fig. 2;][]{Giacalone2003,Giacalone2005}. 

\begin{figure}[h!]
\begin{center}
\includegraphics[width=0.6\textwidth]{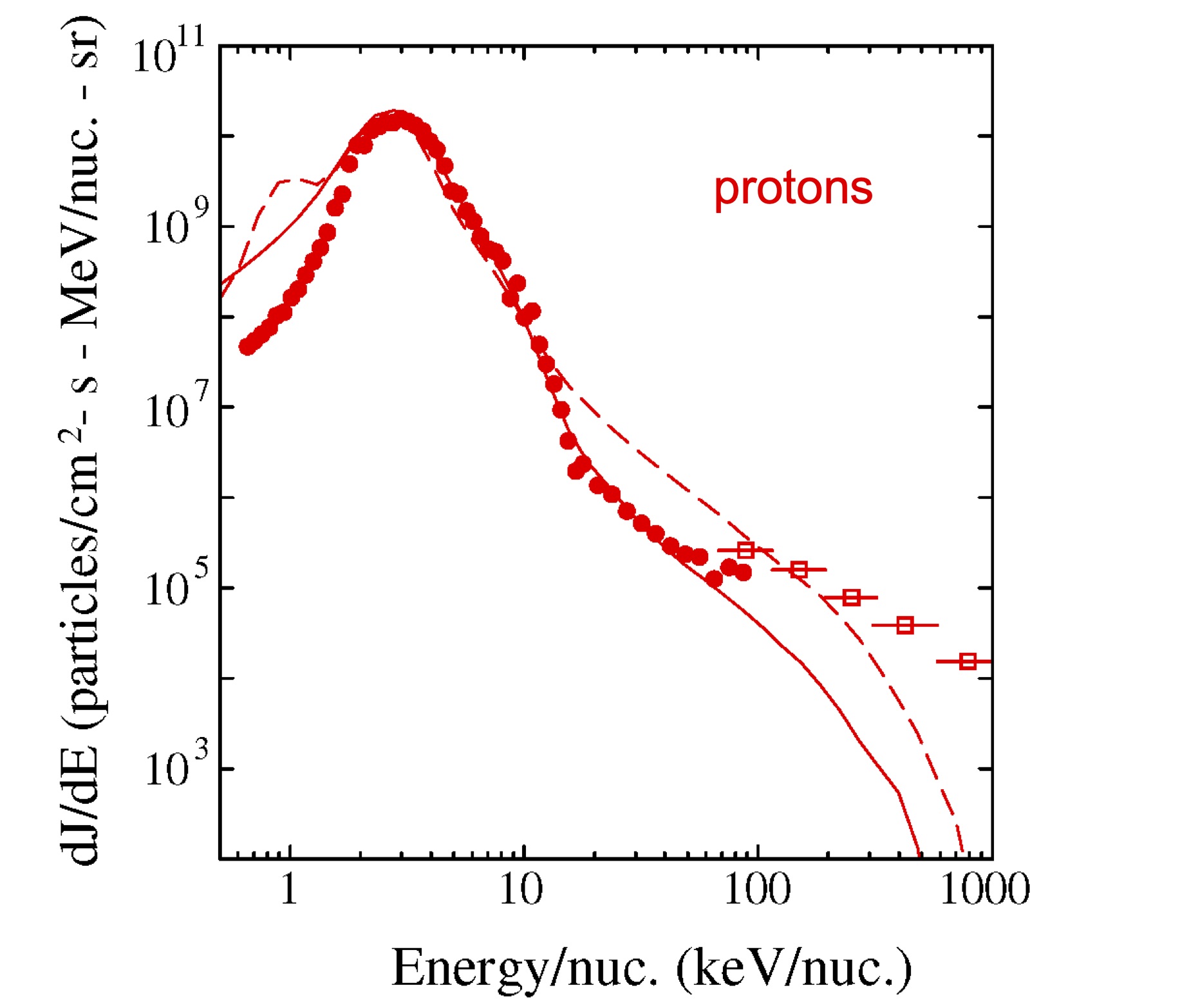}
\end{center}
\caption{Energy spectra at two regions downstream of the shock in the hybrid simulation shown in Figure \ref{fig:proton-distribution} (solid curve $1300<x \omega_{pi}/c < 1500$ and $6300<z \omega_{pi}/c < 7400$) and (dashed curve $1100<x \omega_{pi}/c < 1300$ and $6300<z \omega_{pi}/c < 7400$). Observed downstream spectrum on DOY 118, 2001 by \citet{Lario2019} is also plotted for comparison.}\label{fig:proton-spectrum}
\end{figure}

In Figure \ref{fig:proton-spectrum} we plot the downstream energy spectra from the hybrid simulation mentioned above (c.f. Figure \ref{fig:proton-distribution}) at averaged over two different regions downstream of the shock, and compare it with the downstream spectrum seen by ACE. The regions were identified by finding regions for which the local shock normal angle (at the location of the shock for the particular value of $z$ chosen) was very nearly perpendicular, as in the observations. Note that the simulation results, represented as either solid or dashed curves are both taken downstream of the shock, yet show differences. This means that the spectrum depends on location along the shock in these simulations which suggests this may also be the case in real interplanetary shocks. It is noteworthy that in this simulation, the initial distribution is purely Maxwellian so that the high-energy particles were accelerated (by the shock) from an initially thermal distribution. Moreover, despite the variation along the shock, the intensity of the tail particles is either consistent with, or even larger than the observations, suggesting that the efficiency of accelerating the thermal solar wind in this case is enough to account for the observations. 

Consideration of kinetic physics is required to understand the process by which very low-energy ions, including those from the thermal solar wind, are accelerated by a shock. It is well known that supercritical collisionless shocks require a dissipation mechanism other than that afforded by anomalous resistivity (e.g. \cite{Kennel_etal_1985}). It was established ~40 years ago that such shocks extract a small fraction of the thermal population incident upon a shock, creating a population of ‘specularly reflected’ ions which are reflected in the shock ramp and gyrate in the upstream magnetic field, and are ultimately advected downstream with the plasma flow. This process is well studied at Earth’s bow shock \citep{Gosling1985}.   The reflected particles have an energy that is about the plasma ram energy as measured in the plasma frame of reference. This is much larger than the thermal energy, thus, these particles represent a suprathermal population, albeit with the still rather low energy of the plasma ram energy. It is reasonable to expect that a fraction of these particles can undergo a further reflection at the shock, this time gaining even more energy; and some can be reflected more than once, gaining even more energy. Thus, these particles are the likely source of the high-energy tail in supercritical shocks, including those CME-driven interplanetary shocks that produce the largest SEP events observed, such as the one presented above.

\section{Electron Acceleration at Collisionless Shocks Moving Through a Turbulent Magnetic Field}
\label{sec:Electron}
Collisionless shocks are efficient accelerators for a variety of energetic charged particles observed in the heliosphere. However, the acceleration of electrons at collisionless shocks is generally considered to be more difficult than that of ions. This is primarily due to the fact that the gyroradii of non-relativistic electrons are much smaller compared with that of protons at the same energy (by a factor of $\sqrt{m_p/m_e} \sim 43$), therefore low-energy electrons cannot resonantly interact with the large-scale magnetic turbulence or ion-scale waves close to the shock front. In fact, statistical results have shown that electrons are less commonly accelerated at interplanetary shocks \citep{Lario2003}.

\begin{figure}[h!]
\begin{center}
\includegraphics[width=0.6\textwidth]{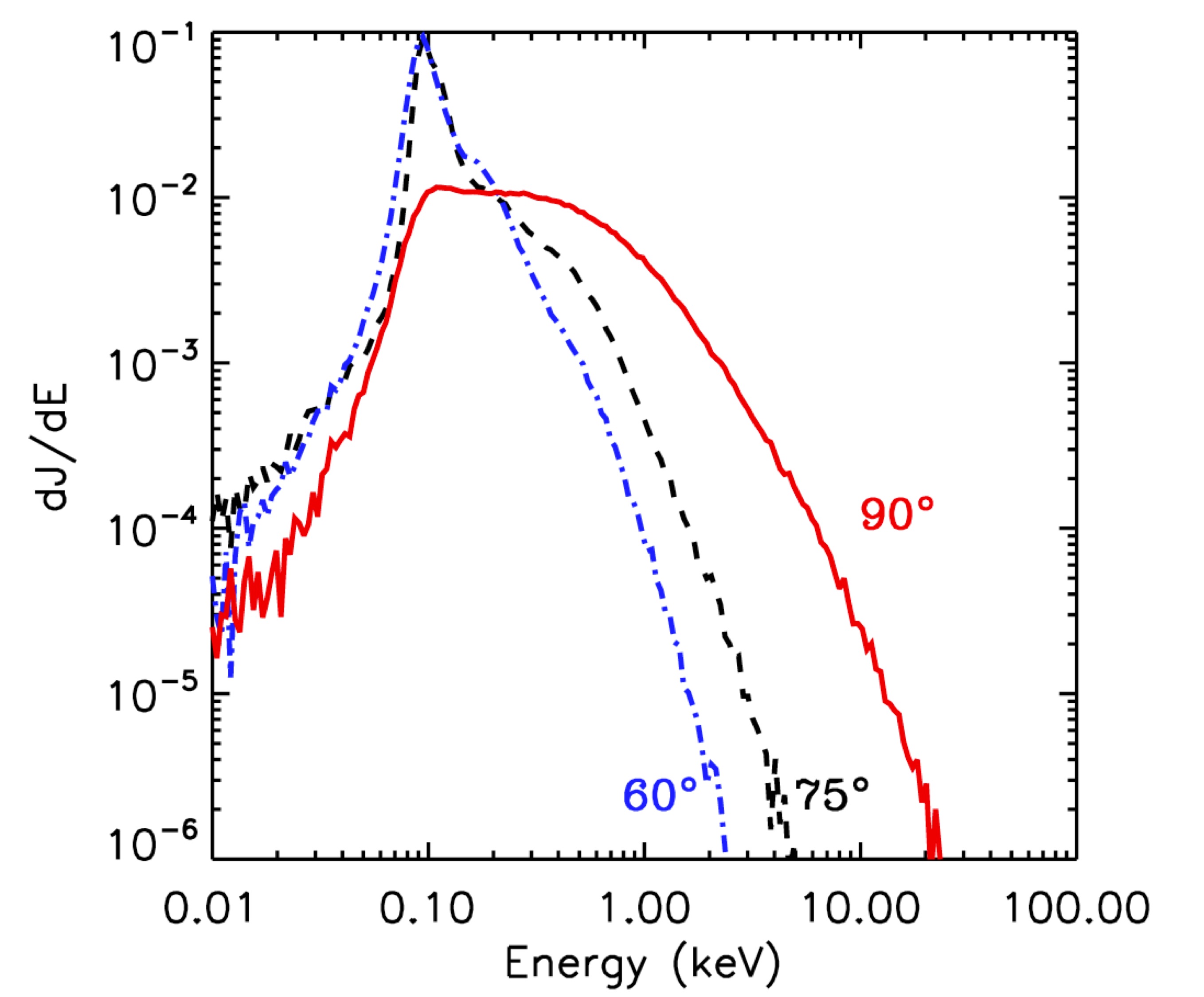}
\end{center}
\caption{Energy flux spectra of electrons for different averaged shock normal angles. Results of hybrid simulations with test-particle electrons adapted from \citet{Guo2010Effect}.}\label{fig:electron-spectrum}
\end{figure}

\subsection{Observations of Electron Acceleration in the Vicinity of Interplanetary Shocks}

At interplanetary shocks, electrons with energy up to $\sim100 $keV have been observed close to quasi-perpendicular shocks with $\theta_{Bn}$ is larger than $45$ degree \citep{Dresing2016,Yang2018,Yang2019}. \citet{Tsurutani1985} reported the observations of energetic electrons associated with
interplanetary shocks showing “spike-like” flux enhancements for energies $> 2$ keV.
The spike events were observed at quasi-perpendicular shocks with $\theta_{Bn} \ge 70^\circ$. Some
shock crossings had no enhancements of energetic electrons that were reported to be
associated with low shock speeds and/or small $\theta_{Bn}$. 
\citet{Simnett2005} presented data which show that energetic electrons are accelerated close to shock front. They also showed some accelerated electrons can escape far upstream of quasi-perpendicular interplanetary shocks. As we will show below, this can be explained by simulations that include large scale fluctuations \citep{Guo2010Effect}. The clear evidence of
electron acceleration at interplanetary shocks by DSA is rare, but an example discussed by \citet{Shimada1999} shows evidence of the importance of whistler
waves (a high frequency wave that can resonantly interact with low-energy electrons) close to a quasi-perpendicular shock (see also \citet{Wilson2012}. This seems to be different from acceleration at Earth's bow shock, where new observations have found whistler waves strongly contribute to the electron acceleration process \citep{Oka2019,Amano2020}, although more observations with high time cadences will definitely be needed to finally resolve this. The observation by Voyager 1 at the
termination shock showed a spike-like enhancement of energetic electrons \citep{Decker2005}. Voyager 2 observed an exponential increase upstream of the termination shock and roughly constant downstream in the heliosheath, similar to what
is predicted from DSA \citep{Decker2008}. Both of the Voyager spacecraft have
observed that electrons are accelerated to at least MeV range, indicating that the
termination shock can efficiently accelerate electrons. 

\begin{figure}[h!]
\begin{center}
\includegraphics[width=\textwidth]{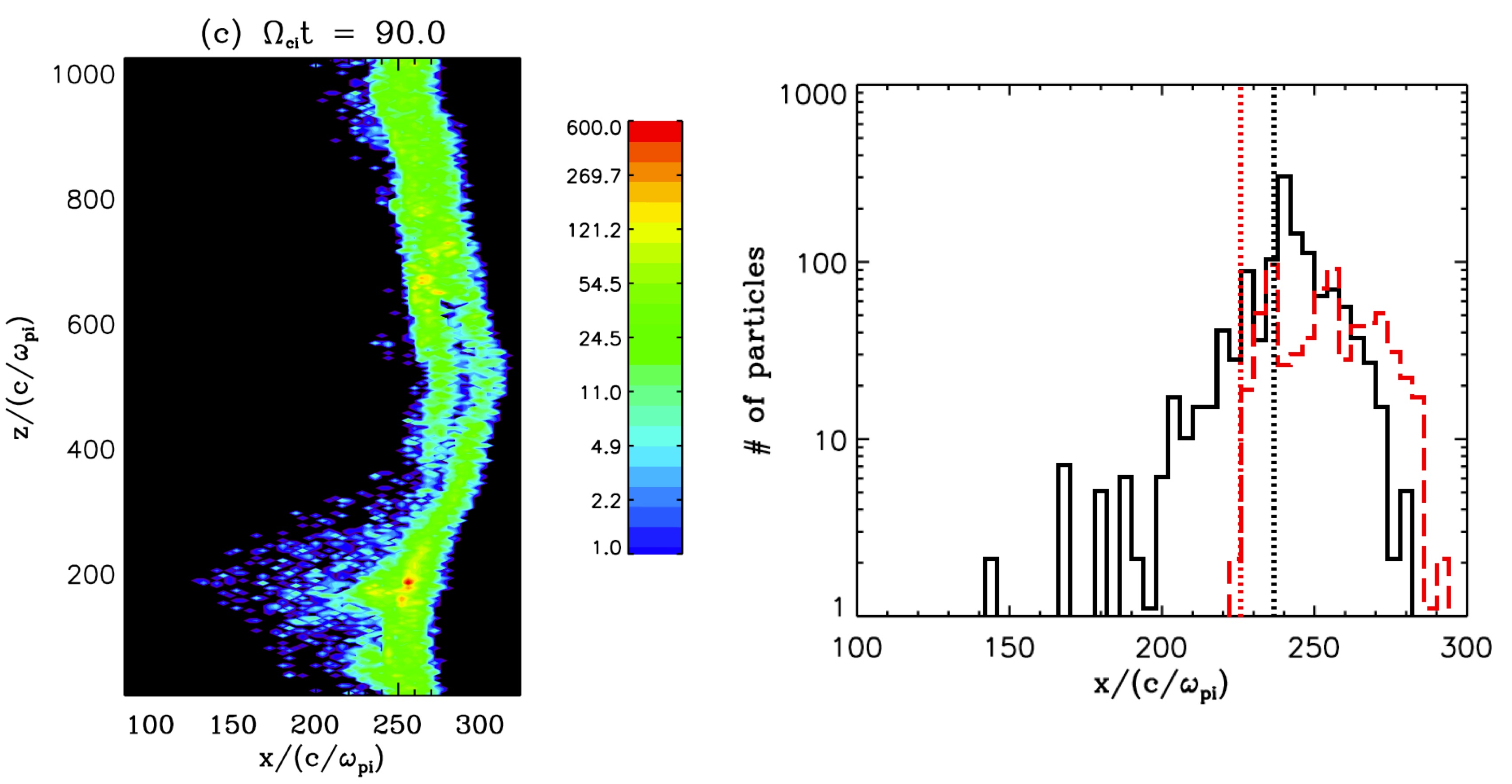}
\end{center}
\caption{Left: Number of energetic electrons with energies $E>10E_0$ in the shock region, where the injected electron energy is $E_0 = 100eV$. Right: two profiles at $z = 200c/\omega_{pi}$ (black) and at $z = 800c/\omega_{pi}$ (red). Adapted from \citet{Guo2010Effect}.}\label{fig:electron}
\end{figure}

\vspace{5pt}
\subsection{One-dimensional Scattering-free Theory}

In order to explain the energization of electrons within the shock layer, \citet{Wu1984} and \citet{Leroy1984} developed analytic models for electron acceleration from thermal energies by adiabatic reflection by a quasi-perpendicular shock (see also \citet{Ball2001}). This process is known as fast-Fermi acceleration, where electrons are reflected by the fast-moving shock along the upstream magnetic field in the de Hoffmann-Teller frame. This theory describes a scatter-free electron acceleration process in a planar, time-steady shock. It obtains a qualitative agreement with observations at Earth’s bow shock in terms of the loss-cone pitch-angle distribution and energy range of accelerated electrons, but cannot explain observed power-low-like downstream energy distribution \citep{Gosling1989}. \citet{Krauss-Varban1989} used the combination of electron test-particle simulation and one-dimensional (1D) hybrid simulation and verified Wu’s basic conclusions. The main energy source of fast-Fermi acceleration comes from the $-\textbf{u} \times \textbf{B}/c$ electric field which is the same as shock drift acceleration (SDA). It can also be demonstrated that fast-Fermi acceleration and SDA are the same process in two different frames of reference \citep{Krauss-Varban1989}. Thus, one would expect electrons to drift in the direction perpendicular to the flow and magnetic field. For a single reflection, the fraction and energies of accelerated particles are limited \citep{Ball2001}. \citet{Holman1983} proposed the basic outline for type II solar radio bursts in which energetic electrons are accelerated through SDA. It is expected that multiple reflections are required in order to explain herringbone structures in type II bursts, where the electrons are accelerated to a fraction of the speed of light.

\begin{figure}[h!]
\begin{center}
\includegraphics[width=0.55\textwidth]{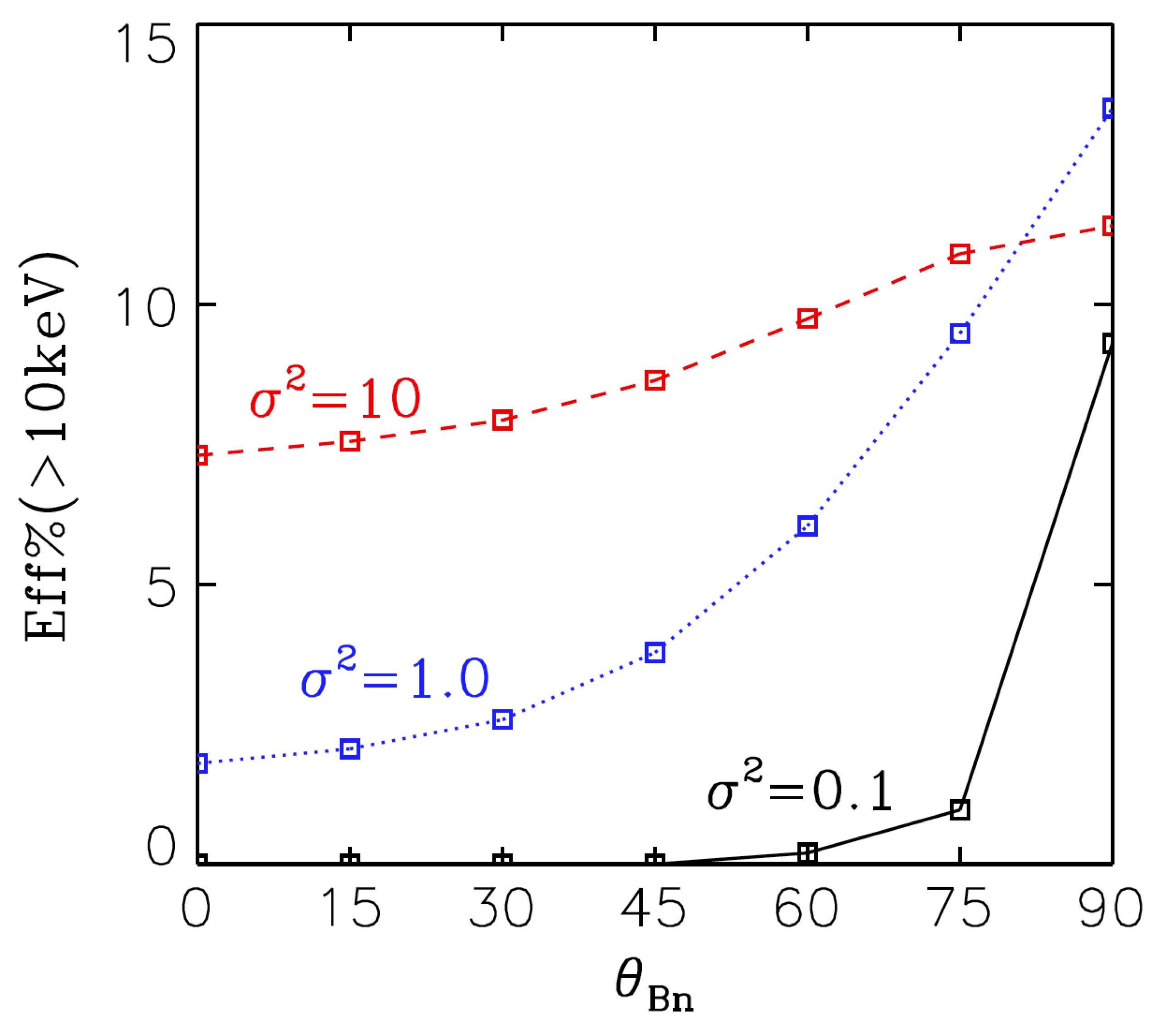}
\end{center}
\caption{Effiency of electron acceleration for various wave variances as a function of shock angle. The efficiency is defined by the fraction of electrons that is accelerated to more than 10 keV at the end of test-particle simulations (adapted from \citet{Guo2015Acceleration}.}\label{fig:effiency}
\end{figure}

\subsection{Effects of Large-scale Magnetic Fluctuations and Shock Ripples}

While the scattering provided by whistler waves \citep{Shimada1999} is one possibility, \citet{Jokipii2007} proposed an attractive mechanism for accelerating electrons to high energy that does not require strong pitch-angle scattering, i.e., conserving the first adiabatic invariant. The idea is that the low-rigidity particles, especially electrons, can move rapidly along meandering magnetic field lines and thus travel back and forth between shock fronts. The particles gain energy from the difference between upstream and downstream flow velocities. \citet{Guo2010Effect,Guo2012Particle,Guo2012Acceleration} have used hybrid simulations combined with test-particle electrons to simulate effects of large-scale, preexisting magnetic fluctuations on electron acceleration at shocks. They found that efficient electron acceleration can happen after considering large-scale pre-existing upstream magnetic turbulence. The turbulent magnetic field leads to field-line meandering that allows the electrons to get accelerated at a shock front multiple times. The rippled surface of the shock front also contributes to the acceleration by mirroring electrons between the ripples. Figure \ref{fig:electron-spectrum} shows electron energy spectra at different averaged shock angles suggesting perpendicular shocks can accelerate electrons more efficiently when upstream magnetic fluctuations exist. 

Energetic particles are often seen upstream of the interplanetary shocks, known as foreshocks.
We note that the spatial distribution of energetic electrons is determined not only by the ripples in the shock front, but also by the global topology of the magnetic field lines. An example is shown in Figure \ref{fig:electron}, which shows the spatial distribution of energetic electrons and profiles of the number of energetic electrons as a function of $x$, for the case of $⟨\theta_{Bn}⟩ = 90^\circ$ in \citet{Guo2010Effect}. The black solid line is the profile at $z = 200c/\omega_{pi}$, and the red dash line shows the profile at $z = 800c/\omega_{pi}$. The corresponding positions of the shock front at each of these values of $z$ are represented using dot lines. At $z = 200c/\omega_{pi}$, it is observed that the energetic electrons travel far upstream. However, the profile at $z = 800c/\omega_{pi}$ shows no significant upstream energetic electron flux. The upstream energetic electron profiles show irregular features similar to in situ observations reported by \citet{Tsurutani1985} and \citet{Simnett2005}. The irregular features are controlled by the global topology of the large-scale turbulent magnetic field lines, along which the accelerated electrons could travel far upstream. Additionally, energetic electron profiles in the $x$-direction generally show “spike” structures close to the shock front, which is usually observed in interplanetary shocks and Earth’s bow shock.  Observation by Voyager 1 at the termination shock and in the heliosheath also shows the evidence of electron spike-like enhancements at the shock front \citep{Decker2005}. The upstream spatial distribution of energetic electrons shows irregular features which depend on both the irregularity in the shock surface and the global topology of magnetic field lines. At first the electrons are accelerated and reflected at the shock front, and then they travel upstream along the magnetic field lines. The electrons could be taken far upstream due to field-line meandering. This result can possibly lead to an interpretation to the complex electron foreshock events  observed to be associated with interplanetary shocks \citep{Bale1999,Pulupa2008}.

The relation between upstream magnetic field turbulence and energetic electrons at shocks has been further studied by \citet{Guo2015Acceleration} through numerical integrating test-particle electron trajectories at a shock that propagates through a prespecified, kinematically defined turbulent magnetic field. Figure \ref{fig:effiency} shows the efficiency of electron acceleration to above $10$ keV (from the initial energy $1$ keV). The acceleration is strongest at perpendicular shocks for all wave variance up to $\sigma= <\delta B^2/B_0^2> = 10$, but quasi-parallel shocks can accelerate electrons when upstream fluctuation amplitude is sufficiently large. This trend in general agrees with observations of electron acceleration in the vicinity of interplanetary shocks \citep{Tsurutani1985,Yang2019}. The acceleration at parallel shock when the wave amplitude is strong enough to explain recent observation at Saturn's bow shock \citep{Masters2013}.

\section{Implications to Variability of Energetic Particles}
\label{sec:Variability}

Large-scale magnetic field line meandering due to magnetic turbulence is ubiquitous in the heliosphere and other astrophysical environments \citep{Jokipii1966,Jokipii1969,Parker1979}.  Including the effects of pre-existing magnetic turbulence is necessary for a complete theory of shock acceleration. 
Although the 1-D steady state DSA solution gives a very elegant description for the acceleration of charged particles at the shock front, nonplanar shock and fluctuating magnetic field effects could play significant role during the acceleration and transport. These may explain many of the observed varabilities of energetic particles at shocks. \citet{Lario2003} have shown that the energetic particle profiles in the vicinity of interplanetary shocks are often not consistent with the 1-D diffusive shock acceleration and many of the profiles are `irregular'. A remarkable example is the in situ observation at the termination shock and in the heliosheath by Voyager 1 \citep{Stone2005}, which found that the intensity of anomalous cosmic rays (ACRs) was not saturated at the place where Voyager 1 crossed the termination shock and kept increasing after entering the heliosheath (although the low energy intensity does peak at the termination shock \citep{Decker2005}), which strongly indicates that a simple planar shock model is inadequate to interpret the acceleration of ACRs. Numerical simulations and analytical studies suggest that possible solutions can be made by considering the temporary and/or spatial variation of magnetic field and the shock surface \citep{Florinski2006,McComas2006,Guo2010Particle,Senanyake2013,Kota2010}.
\citet{Neugebauer2006} have found that the details of energetic particle fluxes in the vicinity of interplanetary shocks are different between ACE and Wind spacecraft for the same interplanetary shocks. The local shock parameters can significantly change, even change from quasi-parallel to quasi-perpendicular, or vice versa. The spatial scale of persistent energetic particle features is about $3\times 10^6 $ km, which is roughly the same as the correlation length of the interplanetary magnetic field \citep{Neugebauer2005}.  

\begin{figure}[h!]
\begin{center}
\includegraphics[width=\textwidth]{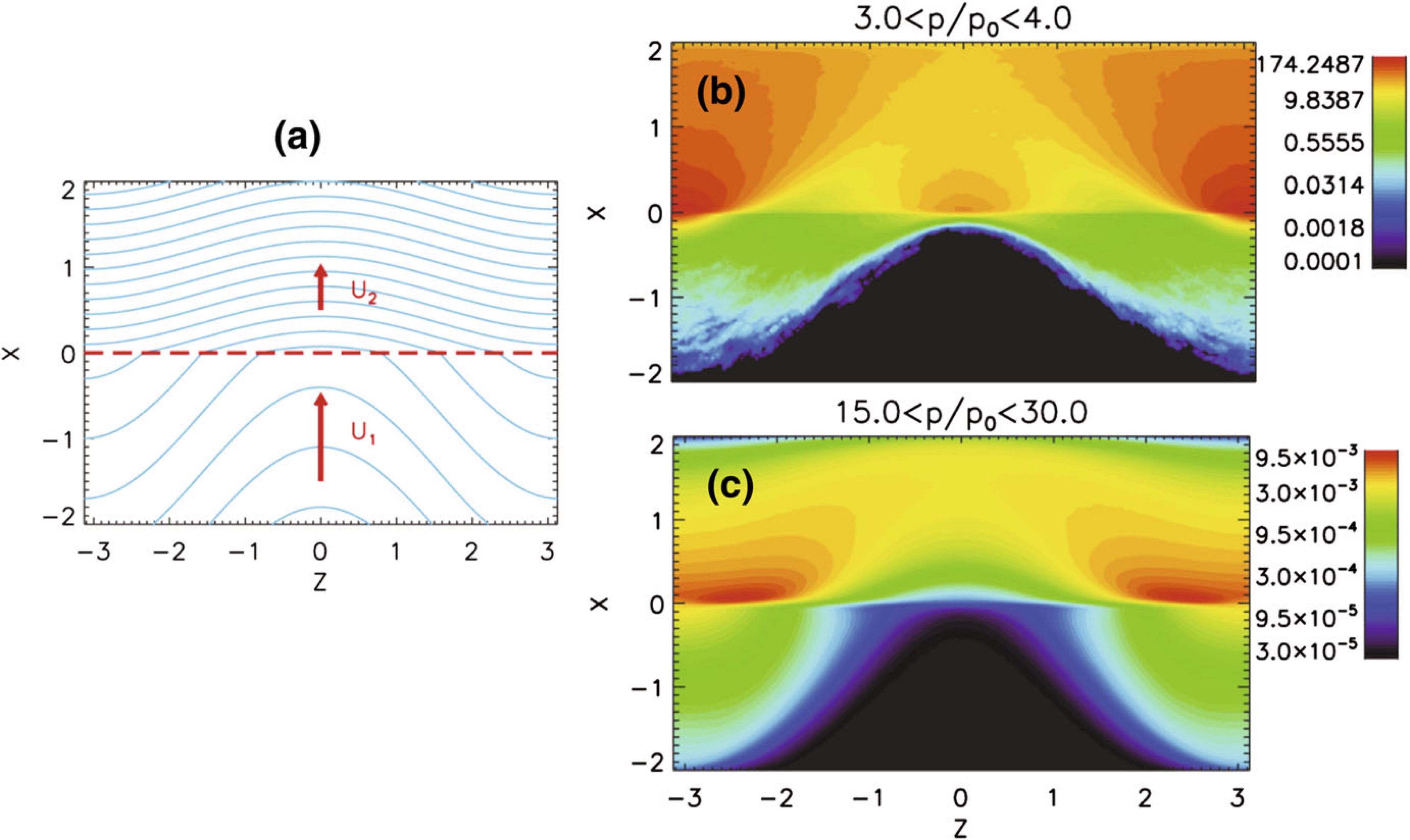}
\end{center}
\caption{a) Shock and magnetic field geometry for an upstream average magnetic field perpendicular to the shock normal. The blue lines represent the magnetic field lines and the red dashed line indicates the shock wave. b) Representation of density contour of accelerated particles, for the energy range $3.0<p/p_0<4.0$, where $p_0$ is the initially injected momentum. c) similar to b) but for $15.0<p/p_0<30.0$ (adapted from \citet{Guo2010Particle}.}\label{fig:acr}
\end{figure}

The acceleration of charged-particles in collisionless shocks has been shown
to be strongly affected by magnetic-field turbulence at different scales \citep{Giacalone2005,Giacalone2008}. The large-scale magnetic field variation
will have important effects on the shock acceleration since the transport of charged
particles is different in the direction parallel and perpendicular to the magnetic
field, as shown in early work \citep{Jokipii1982,Jokipii1987}. Blunt shocks or shocks with
fluctuating fronts \citep{Li2006} that have a similar geometry, are also relevant
to this problem. \citet{Guo2010Particle} analyzed the effect of the large-scale spatial variation
of magnetic field on DSA by considering a simple system that captures the basic
physical ideas (Fig. \ref{fig:acr}). They solve numerically
the Parker transport equation for a shock in the presence of large-scale sinusoidal
magnetic-field variations. 
They demonstrated that the familiar planar-shock results can
be significantly altered as a consequence of large-scale, meandering magnetic lines of
force. Because the perpendicular diffusion coefficient $\kappa_\perp$ is generally much smaller
than the parallel diffusion coefficient $\kappa_\parallel$, energetic charged particles are trapped
and preferentially accelerated along the shock front in regions where the connection
points of magnetic field lines intersecting the shock surface converge, and thus create “hot spots” of accelerated particles \citep[see also][]{Kong2017,Kong2019b}. For regions where the connection points
are separated from each other, the acceleration to high energies will be suppressed.
Furthermore, the particles diffuse away from the “hot spot” regions and modify the
spectra of downstream particle distribution. These results are potentially
important for particle acceleration at shocks propagating in turbulent magnetized
plasmas as well as those that contain large-scale nonplanar structures. For example, in many interplanetary shocks, the peak of the energetic particle intensity is in the downstream rather than at the shock transition \citep{Lario2003}.

\section{Discussion and Summary}
\label{sec:final}

In this review, we have discussed the effects of solar wind turbulence on the propagation of shock waves and on particle acceleration and transport in the vicinity of the shocks. The propagation of interplanetary shocks in the turbulent solar wind leads to rippling shock surfaces on various scales. These structures, along with the upstream and downstream magnetic fluctuations likely contribute to the observed energetic particles close to the interplanetary shocks. We emphasized the role of pre-existing upstream turbulence in enhancing the acceleration of low-energy particles at the shock wave, providing a promising means for solving the well known injection problem, especially at quasi-perpendicular shocks. Upstream turbulence is also important to understand the acceleration of electrons and observations at interplanetary shocks. 
As a remark, solar wind turbulence may be the key for interpreting the variability of energetic particles at interplanetary shocks. This, however, has not been understood in a quantitative way. 
Observations of interplanetary shocks provide a unique opportunity to systematically understand the system where shocks interact with large-scale turbulence and accelerate energetic particles.
Understanding how these processes work is also important to many other space, solar and astrophysical systems \citep{Giacalone2017,Kong2019}.

\section*{Funding}

F. G. and J. G. acknowledge the support in part by NASA LWS program grant 80HQTR21T0005. F. G. also acknowledges F. G. is also supported by
NSF grant AST-1735414 and DOE grant DE-SC0018240. 
L. Z is supported in part by NASA under Grants 80NSSC19K0076, 80NSSC20K1849, and 80NSSC18K0644.

\section*{Acknowledgments}
F. G. thanks Drs. Senbei Du and Dr. Xiaocan Li for discussion on the acceleration of particles and Dr. Mitsuo Oka for discussion on the observations of electron acceleration at interplanetary shocks.



\bibliographystyle{aasjournal} 






\end{document}